\documentclass[aps,prd,twocolumn,nofootinbib,superscriptaddress]{revtex4-2}

\usepackage{amsmath,amssymb,bm}
\usepackage{graphicx}
\usepackage{xcolor}
\usepackage{booktabs}
\usepackage{dcolumn}
\usepackage[colorlinks=true,linkcolor=blue,citecolor=blue,urlcolor=blue]{hyperref}
\usepackage{bm}

\graphicspath{{../figures/}}

\newcommand{\pcrit}{p_{\rm crit}}
\newcommand{\Tn}{T_n}
\newcommand{\Tc}{T_c}

\newcommand{\fhat}{\widehat f_\xi}

\begin{document}

\title{Subcritical bubble prehistory in weak first-order phase transition}

\affiliation{School of Physics, Henan Normal University, Xinxiang 453007, P. R. China}

\author{Guangshang Chen}
\affiliation{Key Laboratory of Particle Physics and Particle Irradiation (MOE), Institute of Frontier and Interdisciplinary Science, Shandong University, Qingdao, 266237, Shandong, China}

\author{Yang Xiao}
\email[]{Corresponding author, xiaoyangphy@gmail.com}
\affiliation{School of Physics, Henan Normal University, Xinxiang 453007, P. R. China}

\author{Jin Min Yang}
\affiliation{School of Physics, Henan Normal University, Xinxiang 453007, P. R. China}
\affiliation{Institute of Theoretical Physics, Chinese Academy of Sciences, Beijing 100190, P. R. China}

\author{Yang Zhang}
\email[]{Corresponding author, zhangyang2025@htu.edu.cn}
\affiliation{School of Physics, Henan Normal University, Xinxiang 453007, P. R. China}

\date{\today}

\begin{abstract}
Standard calculations of cosmological first-order phase transitions usually assume critical bubbles to nucleate on a homogeneous symmetric vacuum background. However, this assumption can fail in weak transitions, where thermal fluctuations trigger subcritical bubbles before the standard nucleation temperature \(T_n\). Motivated by this possibility, we systematically examine whether the homogeneous nucleation background approximation is self-consistent. By evolving the Gelmini-Gleiser subcritical bubble kinetics and comparing it with the standard critical bubble nucleation picture, we identify the parameter regions in which the background becomes apparently mixed. A detailed scan of these regions shows that sizable subcritical volume fractions arise when the two phases are nearly degenerate at \(T_n\), the potential barrier is low, the difference of free energy between the symmetric and broken phases is moderate and the transition strength is weak. Our analysis further yields a simple criterion, \(\log_{10}\hat f_\xi(T_n)\simeq -1.95\), for a percent level subcritical bubble volume fraction. Parameter points above this boundary should be treated as mixed background candidates rather than as ordinary homogeneous bounce points.

\end{abstract}

\maketitle

\section{Introduction}

The electroweak phase transition is an important event in the thermal history of
the early Universe.  In the Standard Model of particle physics, the observed Higgs mass makes this
transition a smooth crossover rather than a first-order phase transition~\cite{Kajantie:1996qd}.
Extensions of the Higgs sector can change this situation, because additional
scalar degrees of freedom may generate a potential barrier and turn the
transition into a first-order one~\cite{Chen:2025ksr,Du:2026qco,Yang:2022quy,Wang:2021ayg,Si:2024vrq,Athron:2023xlk}.  Such a transition leads to vacuum bubble nucleation, whose subsequent
collisions, plasma sound waves, and turbulence may source a stochastic
gravitational wave background within the projected reach of space based
interferometers~\cite{Athron:2023xlk,han2021dark,guo2021phase,yang2023gravitational,vaskonen2017electroweak,beniwal2019gravitational,kang2018strong,chala2018signals, alves2020di, alves2019collider, chao2017gravitational}.  The moving bubble walls can also provide the out-of-equilibrium condition required for
electroweak baryogenesis~\cite{Trodden:1998ym,Cline:2006ts,10.1088/978-1-6817-4457-5,Morrissey:2012db}, while the associated thermal history may affect dark
matter production or primordial black hole formation (PBH) ~\cite{Baker:2018vos,Baker:2019ndr,Baker:2017zwx,Hawking:1982ga,Carr:2020gox,Xiao:2022oaq,Xiao:2023dbb}.
For this reason, first-order electroweak phase transitions provide a direct
link between Higgs sector dynamics and cosmological observables.

\begin{figure}[t]
    \centering
    \includegraphics[width=\columnwidth, height=0.53\columnwidth]{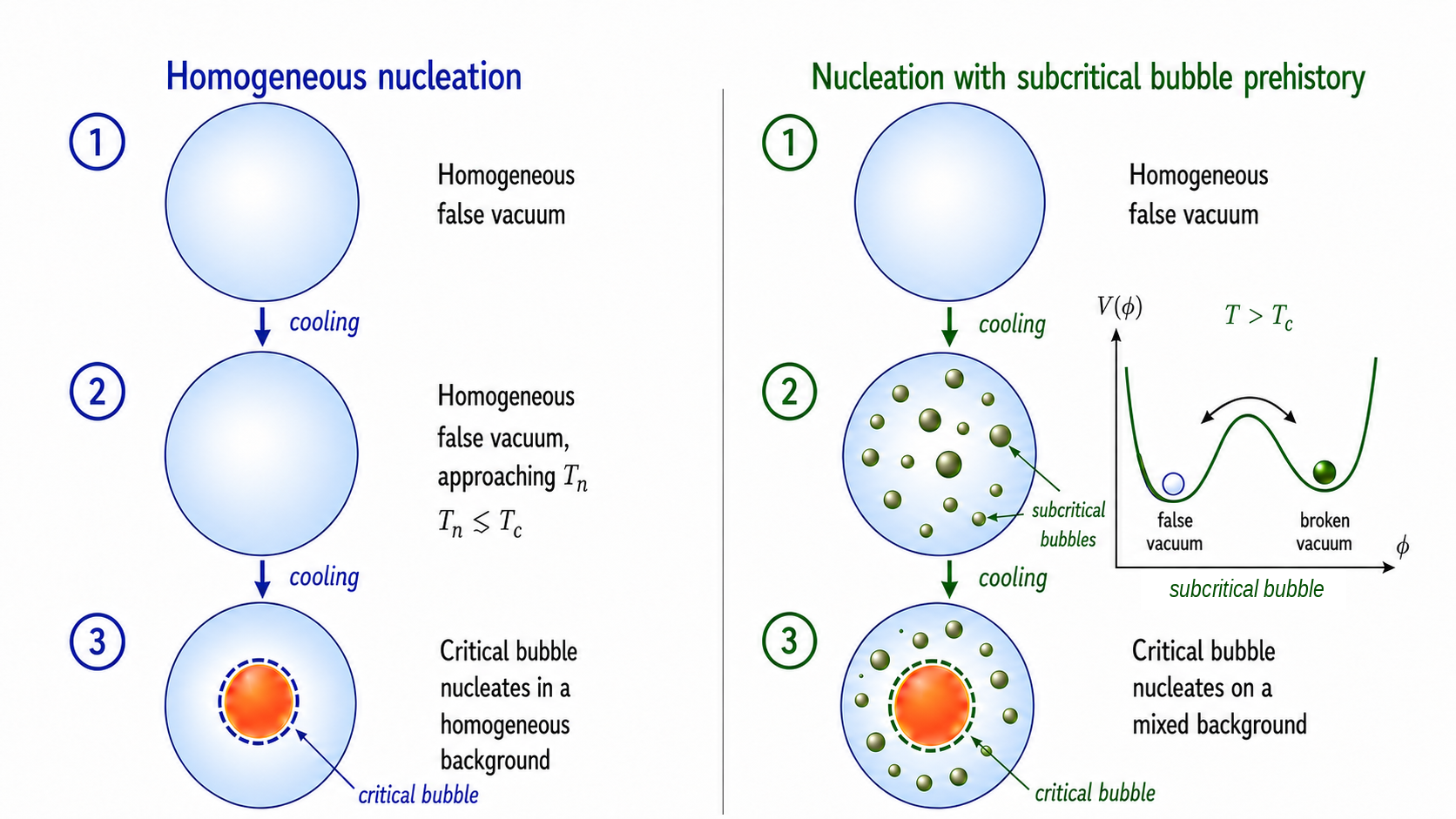}
    \caption{
    A conceptual comparison of the homogeneous nucleation and the nucleation with
    subcritical bubble prehistory.  In the latter case, the critical bubble forms
    on a mixed background rather than in a purely homogeneous false vacuum. This diagram was created with the assistance of ChatGPT.
    }
    \label{fig:concept}
\end{figure}

The standard description of a first-order transition begins with the nucleation
and expansion of critical bubbles.  In Coleman's formulation and most
finite temperature applications, the critical bubble is computed on a
homogeneous symmetric vacuum background~\cite{linde1983decay,Coleman:1977py}.  At finite temperature, however, this background is dynamical.  Thermal fluctuations can push the field
across the barrier toward the broken-phase side of the potential, while reverse
fluctuations can erase these regions before they reach the
critical size~\cite{Gleiser:1991rf,Gleiser:1991zd}.  The pre-nucleation state may therefore contain 
short-lived broken regions that are continuously created and destroyed, as illustrated
in Fig.~\ref{fig:concept}.  Since their sizes are controlled by thermal
correlation scales and remain below the critical radius, these objects are
usually referred to as subcritical bubbles~\cite{Gleiser:1991zd}.

The possible impact of subcritical bubbles was recognized in
Refs.~\cite{Gleiser:1991rf, Anderson:1992uf,Gleiser:1991zd}.  These works argued that, when a
transition is sufficiently weak, thermal fluctuations can populate both phases
and replace the usual homogeneous metastable state by a two-phase emulsion.  The
idea was later developed into a kinetic framework in
Refs.~\cite{Gelmini:1992yj,Shiromizu:1995es,Gleiser:1995er}, where subcritical bubbles are formed and
removed by shrinking, thermal noise, and reverse fluctuations.  Real time
Langevin studies further showed that thermal fluctuations can affect both the
onset of the transition and the subsequent evolution of critical bubbles~\cite{Gelmini:1992yj,Borrill:1994nk}.
These studies therefore indicate that subcritical bubbles can constitute an
important component of phase transition dynamics, rather than a negligible
microscopic correction to the critical bubble picture.  

However. there is no universal criterion for determining how weak a
first-order transition must be before the subcritical population becomes large
enough to challenge the homogeneous background approximation. 
%
It is
important because weak transitions are often not expected to generate sizable
macroscopic signals, as their released energy can be limited.  If subcritical bubbles generate an
appreciable broken phase population before ordinary critical bubble nucleation, a subset of weak transitions may begin from modified initial
conditions, which can affect the subsequent transition history and its
phenomenological consequences.  
As one possible consequence, the additional
small scale structure introduced by subcritical bubbles may modify the
high frequency part of the gravitational wave spectrum~\cite{Bian:2026xdm}.

In this work, we study subcritical bubble prehistory as a consistency test of
the standard homogeneous nucleation framework.  A birth-death kinetic equation
is evolved along a radiation-dominated cooling history, and the resulting true
broken-phase volume fraction is compared with the standard nucleation picture.  A parameter space scan is performed to determine how the subcritical fraction
depends on the transition strength, the free energy splitting, the barrier
height, and the nucleation temperature.  Our analysis gives a practical
criterion for deciding when the homogeneous symmetric vacuum background remains a
self-consistent starting point and when mixed background effects should be
included.

The paper is organized as follows.  Section~\ref{sec:model} introduces the
finite temperature potential and the standard critical bubble calculation.
Section~\ref{sec:kinetics} formulates the subcritical kinetic equation in an
expanding Universe.  Section~\ref{sec:scan} describes the details of the
parameter scan and the numerical inputs.  Section~\ref{sec:results}
presents and discusses the scan results, including representative histories,
the parameter space dependence of the subcritical bubble volume fraction, and the
resulting practical criterion.  It also discusses how a subcritical prehistory
may induce phenomenological corrections to the subsequent phase transition
dynamics.

\section{Theoretical Framework}
\subsection{Homogeneous background critical bubble nucleation}
\label{sec:model}
As a minimal benchmark for weak first order phase transitions, we consider the finite temperature potential of a single scalar field.
\begin{equation}
V(\phi,T)
=
D(T^2-T_0^2)\phi^2
-
ET\phi^3
+
\frac{\lambda}{4}\phi^4 .
\label{eq:potential}
\end{equation}
Its symmetric vacuum is located at
\begin{equation}
\phi_s=0 ,
\end{equation}
At high temperatures, this is the only local minimum.  As the
temperature decreases, two nonzero stationary points can appear, corresponding to the barrier location $\phi_{\rm top}$ and the broken phase $\phi_b$
\begin{equation}
\phi_{ {\rm top},b}(T)
=
\frac{
3ET \mp \sqrt{9E^2T^2-8\lambda D(T^2-T_0^2)}
}{2\lambda}.
\label{eq:phitop_phit}
\end{equation}

The temperature \(T_1\) is the point at which these two nonzero stationary
points first appear.  Equivalently, it is the temperature where the square root
in Eq.~\eqref{eq:phitop_phit} vanishes. This gives
\begin{equation}
T_1^2
=
\frac{8\lambda D}{8\lambda D-9E^2}\,T_0^2 .
\end{equation}
For \(T>T_1\), the nonzero stationary points are absent and the potential
contains only the symmetric minimum.  For \(T<T_1\), the potential contains the
ingredients of a first order transition: a symmetric minimum, a potential barrier, and a broken phase minimum.

The critical temperature \(T_c\) is defined by the degeneracy of the symmetric
and broken phase,
\begin{equation}
V(0,T_c)=V(\phi_c,T_c),
\qquad
V'(\phi_c,T_c)=0 .
\end{equation}
These conditions give
\begin{equation}
\frac{\phi_c}{T_c}
=
\frac{2E}{\lambda}
\equiv r_c,
\qquad
T_c^2
=
\frac{T_0^2}{1-E^2/(\lambda D)} .
\label{eq:Tc}
\end{equation}
Given the finite temperature potential, the critical bubble is obtained from the
finite temperature $O(3)$ bounce equation~\cite{rubakov2009classical,quiros1998finite},
\begin{gather}
    \frac{d^2\phi}{dr^2}
    +
    \frac{2}{r}\frac{d\phi}{dr}
    =
    \frac{\partial V}{\partial\phi},
    \\
    \left. \frac{d\phi}{dr}\right|_{r=0}=0,
    \qquad
    \phi(r\to\infty)=0 ,
    \nonumber
\label{eq:bounceeom}
\end{gather}
where $r$ is the radial coordinate of the bubble.
The boundary condition at spatial infinity assumes that the field approaches the
homogeneous symmetric vacuum. Then the corresponding three dimensional Euclidean action is
\begin{equation}
S_3(T)
=
4\pi\int dr\,r^2
\left[
\frac12\left(\frac{d\phi}{dr}\right)^2
+
V(\phi,T)-V(0,T)
\right].
\label{eq:S3}
\end{equation}
The nucleation rate is exponentially suppressed by this action~\cite{linde1983decay},
\begin{equation}
\Gamma_c(T)
=
A_c(T)\exp[-S_3(T)/T],
\label{eq:critrate}
\end{equation}
where \(A_c\) is the fluctuation prefactor, estimated as  \(A_c\simeq T^4\) in the present work.  More refined treatments of the
prefactor can be found in Ref.~\cite{Ekstedt:2023sqc}.

During radiation dominated period, the
expected number of critical bubbles produced in one Hubble volume is
\begin{equation}
N_c(T)
=
\int_{t_c}^{t(T)} dt'\,
\frac{\Gamma_c(t')}{H^3(t')}
=
\int_T^{T_c}
\frac{dT'}{T'}
\frac{\Gamma_c(T')}{H^4(T')}
 ,
\label{eq:nucleation_integral}
\end{equation}
where \(dT/dt=-HT\) has been used and the nucleation temperature \(T_n\) is defined as 
\begin{equation}
N_c(T_n)=1 .
\end{equation}

To describe the progress of the transition, we also track the volume fraction
converted by critical bubbles.  Under the assumption that critical
bubbles nucleate independently as a Poisson process~\cite{Athron:2023xlk}, the probability that a
given point has been converted to the broken phase is
\begin{equation}
\pcrit(t)=1-\exp[-I_c(t)] .
\label{eq:pcrit}
\end{equation}
Here \(I_c(t)\) is the spacetime volume swept out by bubbles nucleated at earlier
times,
\begin{equation}
I_c(t)=\int_{t_i}^{t}dt'\,
\Gamma_c(t')
\left[\frac{a(t')}{a(t)}\right]^3
\frac{4\pi}{3}R(t,t')^3,
\label{eq:Icrit}
\end{equation}
where the factor \([a(t')/a(t)]^3\) accounts for the dilution of the physical number
density of bubbles due to cosmic expansion, and
\begin{equation}
R(t,t')=a(t)\int_{t'}^{t}\frac{v_w\,dt''}{a(t'')}
\label{eq:growth}
\end{equation}
is the physical radius at time \(t\) of a bubble nucleated at time \(t'\).  In
the numerical analysis, we take the bubble wall velocity \(v_w=1\). Modifying this quantity has little effect on the final result.

The transition strength is quantified by the ratio of vacuum energy released to the radiation energy density
\begin{equation}
\alpha(T)
=
\frac{\Delta \rho(T)}{\rho_{\rm rad}(T)} ,
\qquad
\rho_{\rm rad}(T)
=
\frac{\pi^2 g_* T^4}{30},
\end{equation}
where
\begin{equation}
\Delta\rho(T)
=
V(0,T)-V(\phi_b,T)
-
T\left[
\partial_T V(0,T)-\partial_T V(\phi_b,T)
\right],
\end{equation}
$g_*$ is the number of relativistic degrees of freedom and \(\alpha_n\equiv \alpha(T_n)\) denotes the transition strength evaluated at
the nucleation temperature,

\subsection{Subcritical bubble prehistory as a background consistency test}
\label{sec:kinetics}
The standard bounce solution calculation assumes that the symmetric vacuum remains spatially homogeneous until critical bubbles nucleate.  Subcritical bubbles challenge the validity of this assumption before \(T_n\).  Following the convention of
Ref.~\cite{Gelmini:1992yj}, we denote \(n(R,t)\) as the number density of subcritical bubbles with fixed radius \(R\) at time \(t\). The kinetic equation in radius space then reads
\begin{equation}
\frac{\partial n}{\partial t}+3Hn
=-\frac{\partial n}{\partial R}\frac{dR}{dt}
+\frac{V_s}{V_{\rm tot}}\Gamma_{s\to b}^{\rm sub}
-\frac{V_b^{(R)}}{V_{\rm tot}}\Gamma_{b\to s}^{\rm sub}
-\frac{V_b^{(R)}}{V_{\rm tot}}\Gamma_{\rm TN}^{\rm sub},
\label{eq:GG}
\end{equation}
where \(V_{\rm tot}\) is the total physical volume, \(V_s\) is the symmetric vacuum portion, and \(V_b^{(R)}\) denotes the broken vacuum volume associated with the fixed radius \(R\) component of the subcritical population. The terms on the right-hand side describe shrinking in radius space, thermal production in the remaining symmetric volume, reverse subcritical nucleation, and erasure by thermal noise.  The term $3Hn$ accounts for the dilution caused by cosmic expansion.  This contribution is absent in the original kinetic treatment, but it is required here because we follow the subcritical bubble population along a cosmological cooling history.

We next specify the rates in Eq.~\eqref{eq:GG}.  For the \(s\to b\) source, we
adopt a Gaussian ansatz for the field profile,
\[
\phi_{b}^{\rm sub}(r,T)=\phi_b(T)\exp[-r^2/R^2].
\]
Substituting this ansatz into the potential gives the free energy
\begin{equation}
F_b(R,T)=A_\nabla(T)R+B_b(T)R^3,
\label{eq:Fplus}
\end{equation}
where
\begin{equation}
A_\nabla=\frac{3\pi^{3/2}}{4\sqrt{2}}\phi_b^2,
\label{eq:Agrad}
\end{equation}
and
\begin{equation}
B_b=\pi^{3/2}\left[
\frac{D(T^2-T_0^2)\phi_b^2}{2^{3/2}}
-\frac{ET\phi_b^3}{3^{3/2}}
+\frac{\lambda\phi_b^4}{4^{5/2}}
\right].
\label{eq:Bplus}
\end{equation}
Then the source rate can be written as 
\begin{equation}
\begin{aligned}
\Gamma_{s\to b}^{\rm sub}(R,T)
&=w_b(T)A_{\rm sc}T^4
\exp[-F_b(R,T)/T]  \\
&\quad\times \Theta(R-\xi_f)\Theta(R_c-R).
\end{aligned}
\label{eq:subsource}
\end{equation}
This term is evaluated only for fluctuations with radii larger than the correlation length, since smaller fluctuations cannot be reliably treated as subcritical bubbles. We therefore impose constraints \(R>\xi_s\), with 
\(\xi_s(T)=[V''(0,T)]^{-1/2}\) denoting the correlation length in the symmetric vacuum. We also require \(R<R_c\), where \(R_c=\sqrt{-A_\nabla/(3B_b)}\) for \(B_b<0\), and \(R_c=\infty\) otherwise. 
Because a thermal fluctuation is characterized by both its size $R$ and its amplitude, fixing the central field value as $\phi_b(T)$ in the Gaussian ansatz effectively treats every resolved fluctuation of radius $R$ as a broken-phase patch. This would overestimate the source near $T_1$, where the broken-side interval $\phi_b-\phi_{\rm top}$ only just opened and remains small. We therefore include a phenomenological factor to account for the fraction of fluctuations that actually cross the barrier
\begin{equation}
w_b(T)=
\left[
\frac{\phi_b(T)-\phi_{\rm top}(T)}{\phi_b(T)}
\right]^2
\Theta(T_1-T)
\label{eq:fieldrange}
\end{equation}
This phenomenological factor
estimates the fraction of amplitude fluctuations that cross the barrier, makes
the source vanish smoothly as the broken side interval closes at \(T_1\), and
approaches order unity once a sizable broken side region is available.

The reverse process is described by a symmetric fluctuation inside a
subcritical bubble.  With the profile
\(\phi_{s}^{\rm sub}(r,T)=\phi_b(T)[1-\exp(-r^2/R^2)]\), its free energy is written as
\begin{equation}
F_s(R,T)=A_\nabla(T)R+B_s(T)R^3,
\label{eq:Fzero}
\end{equation}
where
\begin{align}
B_s \simeq \pi^{3/2}\Big[
&-1.646\,D(T^2-T_0^2)\phi_b^2
+2.132\,ET\phi_b^3
\nonumber\\
&-0.631\,\lambda\phi_b^4
\Big].
\label{eq:Bzero}
\end{align}
The corresponding reverse rate is also restricted to resolved subcritical symmetric
fluctuations,
\begin{equation}
\Gamma^{\rm sub}_{b\to s}(R,T)
=A_{\rm sc}T^4 e^{-F_s(R,T)/T}
\Theta(R-\xi_b)\Theta(R_{c,s}-R),
\label{eq:mainreverserate}
\end{equation}
where \(\xi_b=[V''(\phi_b,T)]^{-1/2}\) is the correlation length in the broken phase. The upper cutoff is
\(R_{c,s}=\sqrt{-A_\nabla/(3B_s)}\) when \(B_s<0\).  If \(B_s\ge0\), no finite
reverse critical radius is imposed.

Following Ref.~\cite{Gelmini:1992yj}, we estimate the thermal noise erasure
rate as
\begin{equation}
\Gamma_{\rm TN}^{\rm sub}(R,T)=\frac{aT}{4\pi R^3/3}.
\label{eq:mainthermalnoise}
\end{equation}
The remaining ingredient in Eq.~\eqref{eq:GG} is the volume factors. In the dilute limit, and in
the notation of Ref.~\cite{Gelmini:1992yj}, the fixed radius volume
fraction is approximated by
\begin{equation}
\frac{V_b^{(R)}}{V_{\rm tot}}
\simeq \frac{4\pi R^3}{3}n(R,t) .
\label{eq:binVplus}
\end{equation}
This approximation is reliable only at the early stage of subcritical bubble production.
At later times it can fail because collisions, mergers, and other nonlinear
processes invalidate the dilute approximation. 

With the definition of \(n(R,t)\), the dilute broken-phase volume fraction carried by the subcritical bubbles is
\begin{equation}
f_b(T)=
\int dR\,
\frac{4\pi R^3}{3}n(R,t).
\label{eq:fplus}
\end{equation}
This expression sums the geometric volume associated with the radius resolved
number density.  It is adequate only when \(f_b\ll1\), where overlap among
subcritical bubbles can be neglected.  When the occupied volume becomes
sizable, however, the linear expression does not account for random overlap and
can no longer be interpreted as a probability.  To avoid this limitation while
keeping the calculation fast, we use a Poisson empty volume probability.
Analogous to the critical bubble probability in Eq.~\eqref{eq:pcrit}, the
broken phase volume fraction generated by subcritical bubbles is written as
\begin{equation}
p_{b}^{\rm sub}(T)=1-e^{-f_b(T)}.
\label{eq:psub}
\end{equation}
The exponential form in Eq.~\eqref{eq:psub} accounts for random overlap among
subcritical bubbles.  At the early stage, where \(f_b\ll1\), one may expand
\(p_{b}^{\rm sub}=1-e^{-f_b}\simeq f_b\).  In this dilute limit, \(f_b\) itself
is the broken-phase volume fraction carried by subcritical bubbles.
The symmetric vacuum fraction entering the source term is then
\begin{equation}
\frac{V_s}{V_{\rm tot}}\simeq p_s^{\rm sub}(T)\equiv 1-p^{\rm sub}_{b}(T)=e^{-f_b(T)} .
\label{eq:falsepoisson}
\end{equation}

Combining these ingredients and taking \(dR/dt=-v\) gives the final equation
\begin{equation}
\frac{\partial n}{\partial t}+3Hn
=v\frac{\partial n}{\partial R}
+p_s^{\rm sub}\Gamma_{s\to b}^{\rm sub}
-aTn
-\frac{4\pi R^3}{3}\Gamma^{\rm sub}_{b\to s}n .
\label{eq:combined}
\end{equation}
Its solution value at the nucleation temperature is
then used to test whether the homogeneous symmetric vacuum background is a self consistent starting assumption.

\section{Scan strategy}
\label{sec:scan}
To determine when subcritical prehistory can modify the standard
nucleation description, we perform a parameter scan in the following
region,
\begin{equation}
D\in[0.05,0.35],\qquad
\lambda\in[0.03,0.30],\qquad
r_c\in[0.035,1.0],
\label{eq:scan}
\end{equation}
For each point these inputs are mapped to
\begin{equation}
E=\frac{\lambda r_c}{2},\qquad
T_0=v_0\sqrt{\frac{\lambda}{2D}},\qquad v_0=246~{\rm GeV}.
\label{eq:param}
\end{equation}
We keep only points satisfying \(D>0\), \(\lambda>0\), \(E>0\),
\(E\lesssim0.10\), and \(E^2/(\lambda D)<0.5\) to achieve a weak first order phase transition.  For each accepted point, we
compute \(S_3(T)/T\) with the \texttt{CosmoTransitions}~\cite{Wainwright:2011kj} and determine \(T_n\) from
Eq.~\eqref{eq:nucleation_integral}. Only points with a finite nucleation integral crossing,
and finite \(S_3(T_n)/T_n\) are retained for the analysis.  The evolution range is
\begin{equation}
T_{\rm start}=1.2\Tc,\qquad
T_{\rm end}=\max(0.8\Tn,1.001T_0),
\label{eq:Trange}
\end{equation}
Unless stated otherwise, we use
\(g_*=106.75\), \(A_{\rm sc}=1\), \(a=0.5\), \(v=1\), and \(v_w=1\).  Details of the numerical implementation, including the finite difference update scheme and numerical consistency checks, are summarized in Appendices~\ref{app:numerics} and~\ref{app:validation}.

\section{Results and Discussions}
\label{sec:results}
The scan described in Sec.~\ref{sec:scan} yields 3245 accepted points in the parameter space.
For each point, we solve Eq.~\eqref{eq:combined} and evaluate
\(p_b^{\rm sub}(T_n)\), the subcritical bubble volume fraction
measured at the nucleation temperature. This quantity tests whether the homogeneous symmetric vacuum assumption remains valid. To summarize the scan, we classify the points using three reference values,
\begin{equation}
p_b^{\rm sub}(T_n)
=\left\{10^{-3},\,10^{-2},\,10^{-1}\right\}. \notag
\label{eq:criterion}
\end{equation}
The points with
\(p_b^{\rm sub}(T_n)<10^{-3}\) are labeled as safe critical; those with
\(10^{-3}<p_b^{\rm sub}(T_n)<10^{-2}\) are labeled as subcritical correction;
those with \(10^{-2}<p_b^{\rm sub}(T_n)<10^{-1}\) are labeled as prehistory
relevant; and those with \(p_b^{\rm sub}(T_n)>10^{-1}\) are labeled as dilute
breakdown. 

With this definition, the full kinetic setup gives 1787 safe critical points,
160 subcritical correction points, 352 prehistory relevant points, and 946
dilute breakdown points.  In the first two classes, the homogeneous
assumption remains self-consistent, albeit with possible small corrections.  In the
third class, the critical bubble will nucleate in a mixed local environment.  The fourth class marks the regime where the dilute approximation used for
\(V_b^{(R)}/V_{\rm tot}\) in Eq.~\eqref{eq:binVplus} is no longer valid.
Thus, the large population of dilute-breakdown points should not be interpreted as evidence that subcritical bubbles reach percolation.
Rather, it shows that, over a large part of the parameter scan, the dilute approximation can fail before the temperature reaches
\(T_n\).  

\begin{figure*}
\includegraphics[width=\textwidth]{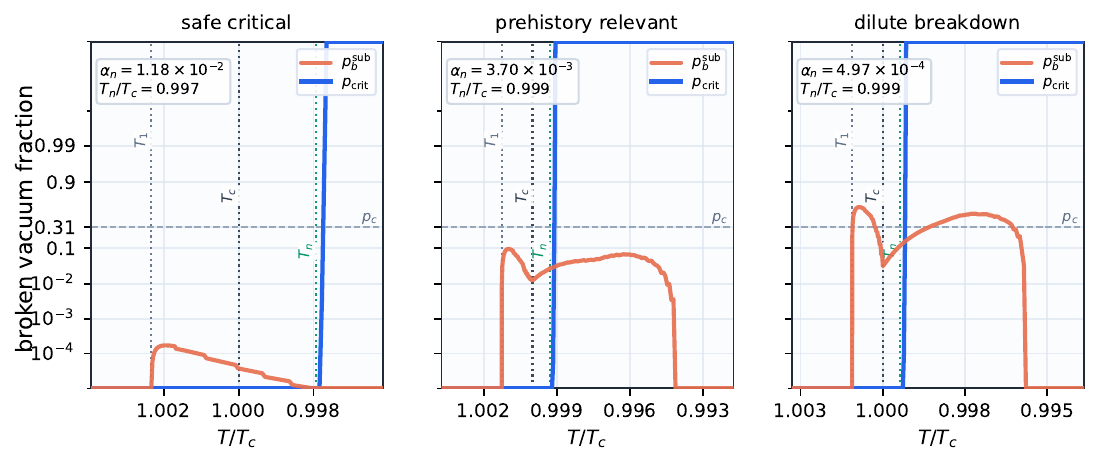}
\caption{\label{fig:rep}
The pre-transition histories for different benchmark points.  The orange curve gives the
subcritical broken vacuum volume fraction \(p^{\rm sub}_b=1-e^{-f_b}\),
while the blue curve gives the critical bubble probability \(p_{\rm crit}\).}
\end{figure*}

The representative histories in Fig.~\ref{fig:rep} show how the above classes
arise dynamically.  For a safe critical point, \(p_{\rm crit}\) rises while the
subcritical bubble volume fraction remains negligible.  For a prehistory relevant
point, subcritical bubbles are already present before the temperature reaches
\(T_n\), while the dilute approximation for \(V_b^{(R)}/V_{\rm tot}\) remains adequate.
For a dilute breakdown point, the subcritical bubble volume fraction becomes so large
that neither the homogeneous background nor the dilute approximation can be trusted as a complete description.

A common feature of the three representative histories in Fig.~\ref{fig:rep} is the sharp
turn-on of the subcritical bubble volume fraction at the appearance temperature $T_1$.
For $T>T_1$, the broken phase is absent, and the subcritical source is
kinematically forbidden. At $T_1$, the broken phase and the potential barrier first appear with $\phi_b=\phi_{\rm top}$, so the amplitude factor $w_b$ still
vanishes. Immediately below $T_1$, however, the interval
$\phi_b-\phi_{\rm top}$ opens and the source becomes nonzero. Because the
microscopic formation and erasure time scales are much shorter than the Hubble cooling time, the subcritical bubble volume fraction rapidly relaxes to its instantaneous kinetic balance. This produces the steep initial rise of $p_b^{\rm sub}$ seen in all three panels.

\begin{figure*}
\includegraphics[width=\textwidth]{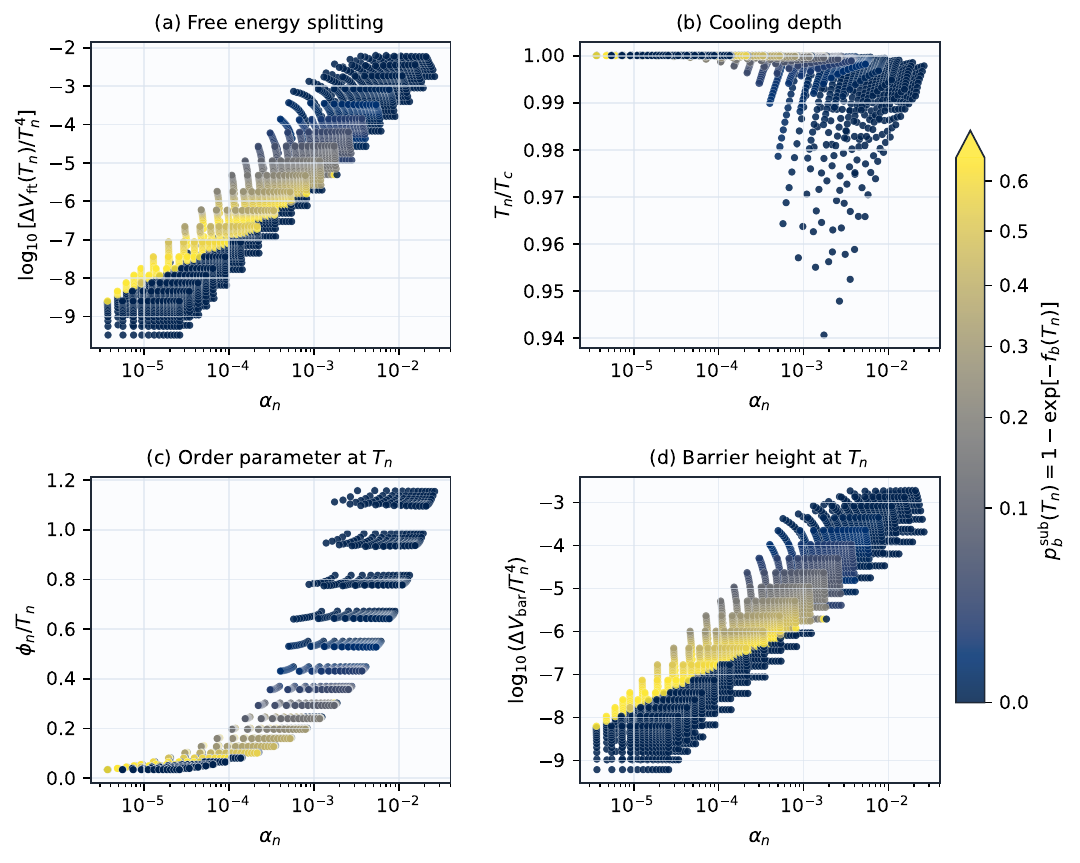}
\caption{\label{fig:scan}
The relationship between transition properties and the subcritical broken vacuum volume fraction.  Each
point represents one accepted scan point and is colored by
\(p^{\rm sub}_b(T_n)\).  Panel (a) compares the
free energy splitting
\(\Delta V_{\rm ft}(T_n)\), normalized by \(T_n^4\),
with \(\alpha_n\).  Panel (b) shows the cooling depth \(T_n/T_c\).  Panel (c)
shows the order parameter value at nucleation, \(\phi_n/T_n\) with
\(\phi_n=\phi_b(T_n)\).  Panel (d) compares the dimensionless barrier height
\(\Delta V_{\rm bar}(T_n)/T_n^4\) with \(\alpha_n\), where
\(\Delta V_{\rm bar}(T_n)\). }
\end{figure*}

The related physical factors of the subcritical bubble volume fraction are shown in
Fig.~\ref{fig:scan}.  Each accepted scan point is colored by $p^{\rm sub}_b(T_n)$
so the gradual growth of the subcritical bubble volume fraction is visible.  We use
\begin{equation}
\Delta V_{\rm ft}(T_n)
\equiv V(\phi_s,T_n)-V(\phi_b,T_n),
\end{equation}
for the finite temperature free energy splitting between the two phases, and
\begin{equation}
\Delta V_{\rm bar}(T_n)
\equiv V(\phi_{\rm top},T_n)-V(\phi_s,T_n),
\end{equation}
for the barrier height above the false vacuum.  The broken phase field value at
nucleation is denoted by \(\phi_n\equiv\phi_b(T_n)\).

Fig.~\ref{fig:scan} shows that the subcritical bubble prehistory is controlled
by the overall structure of the finite-temperature potential.  Panel (a) shows
that large subcritical bubble volume fractions arise when the two phases remain nearly
degenerate at \(T_n\), for which the free energy cost of occupying the broken side is small.  Panel (b) presents the corresponding cooling history, where the largest fractions are concentrated near \(\Tn/\Tc\simeq1\).

Panels (c) and (d) further identify the other ingredients that make this
behavior possible.  A smaller \(\phi_n/T_n\) indicates that the broken phase lies closer to the symmetric point at \(T_n\), reducing both the field excursion and the gradient energy cost of forming a subcritical bubble.  A smaller \(\Delta V_{\rm bar}(T_n)/T_n^4\) weakens the Boltzmann suppression in the formation rate. In each panel, however, the plotted quantity provides only
a partial ordering of the final subcritical bubble volume fraction.  No single
quantity fully determines \(p_b^{\rm sub}(T_n)\).  Instead, the formation and
survival of subcritical bubbles are governed by the combined effects of the finite temperature potential, the cooling history, and the microscopic dynamics.

These trends motivate a fast criterion, which can be obtained by reducing the kinetic equation to a single representative bin equation.  Starting from
Eq.~\eqref{eq:combined}, we first integrate over the bubble radius.  The total
number density of subcritical bubbles is defined as
\begin{equation}
\mu(t)=\int_0^\infty dR\, n(R,t).
\end{equation}
This gives
\begin{equation}
\frac{d\mu}{dt}
=
\int_0^\infty dR\,v\,\frac{\partial n}{\partial R}
+
S_{\rm tot}(T)
-
\int_0^\infty dR\,K(R,T)n(R,t),
\label{eq:mu_integrated}
\end{equation}
where
\begin{equation}
S_{\rm tot}(T)
=
\int_0^\infty dR\,
p_s^{\rm sub}(T)\Gamma_{f\to b}^{\rm sub}(R,T),
\end{equation}
and
\begin{equation}
K(R,T)
=
\frac{4\pi R^3}{3}
\left[
\Gamma_{b\to f}^{\rm sub}(R,T)
+
\Gamma_{\rm TN}^{\rm sub}(R,T)
\right]
+3H .
\label{eq:K_fast}
\end{equation}
The first term on the right-hand side of Eq.~\eqref{eq:mu_integrated} is a
boundary term in radius space.  For a constant shrinkage speed,
\begin{equation}
\int_0^\infty dR\,v\,\frac{\partial n}{\partial R}
=
v\,[n(\infty,t)-n(0,t)] ,
\end{equation}
which vanishes for the boundary conditions relevant to the 
distribution of subcritical bubble.

The remaining radius integral can be approximated by evaluating the integrand at
a representative radius \(\xi \sim \xi_s = [V''(0,T)]^{-1/2}\).  In this single-bin approximation,
\begin{equation}
\int_0^\infty dR\,K(R,T)n(R,t)
\simeq K(\xi_s,T)\,\mu(t), 
\end{equation}
and
\begin{equation}
    \begin{aligned}
        S_{\rm tot}(T)&\simeq p_s^{\rm sub}(T) S^{*}(\xi_s, T)  \\
        &= p_s^{\rm sub}(T)\,
        \Gamma_{f\to b}^{\rm sub}(\xi_s,T)\,
        \Delta R_\xi(\xi_s, T),
    \end{aligned}
\end{equation}
where
\begin{equation}
\Delta R_\xi(\xi, T)
\equiv
\frac{T}{F_b'(\xi,T)}
\left[
1-
\exp\left(
-\frac{F_b'(\xi,T)}{T}(R_c-\xi)
\right)
\right].
\end{equation}
Whereas,the integrated kinetic equation reduces to
\begin{equation}
\frac{d\mu}{dt}
=
S_{\rm tot}(T)
-
K(\xi_s,T)\mu .
\label{eq:single_bin_mu}
\end{equation}
If the microscopic relaxation time is shorter than the Hubble cooling time, the
system rapidly reaches a quasi-stationary state with \(d\mu/dt\simeq0\).
Eq.~\eqref{eq:single_bin_mu} then gives
\begin{equation}
\mu_{\rm qs}(T)
\simeq
\frac{S_{\rm tot}(\xi_s,T)}
{K(\xi_s,T)}.
\end{equation}
At the early stage of subcritical bubble production, the broken phase volume fraction is negligible, and hence \(p_{s}^{\rm sub}\simeq1\).  Therefore,
\begin{equation}
\mu_{\rm qs}(T)
\simeq
\frac{S^{*}(\xi_s,T)}
{K(\xi_s,T)}.
\end{equation}

Our goal is to estimate the volume occupied by subcritical
bubbles,
\begin{equation}
f_b^{\rm sub}(T)
\equiv
\frac{V_b}{V_{\rm tot}}
\simeq
\int_0^\infty dR\,\frac{4\pi R^3}{3}n(R,t),
\end{equation}
which provides the necessary input for estimating \(p^{\rm sub}_b\).  Applying the same single-bin approximation gives the following
\begin{equation}
f_b^{\rm sub}(T)
\simeq
\frac{4\pi\xi_s^3}{3}
\frac{S^{*}(\xi_s,T)}
{K(\xi_s,T)}.
\label{eq:single_bin_fb}
\end{equation}
This fast criterion contains the same ingredients that control the full scan: the volume of a correlation-sized subcritical bubble, the Boltzmann suppression from the free energy, and the lifetime set by erasure.

\begin{figure}
\includegraphics[width=\columnwidth]{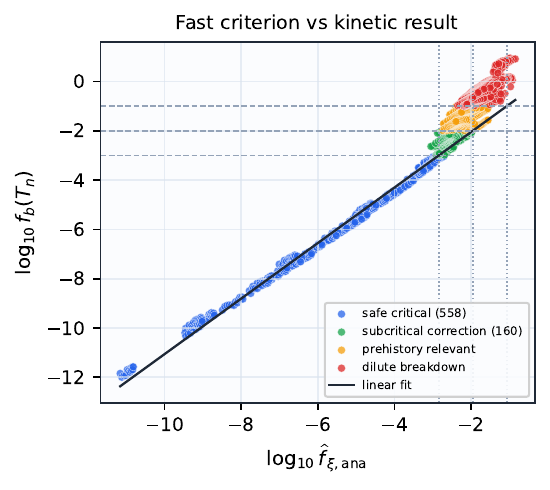}
\caption{\label{fig:fastcrit} Comparison between the fast criterion and the full kinetic result for the fitting sample.  The
displayed points have finite \(\fhat(T_n)\) and \(f_b(T_n)>10^{-12}\).  The
dashed lines show the \(10^{-3}\), \(10^{-2}\), and \(10^{-1}\) volume fraction
thresholds used in Eq.~\eqref{eq:criterion}.  The numbers in the legend give
the selected sample size in each class.}
\end{figure}

To quantify how the theoretical estimate differs from the full numerical
result, we fit the relation between \(\widehat f_\xi(T_n)\) and the kinetic
result in the logarithmic space.  Since the fast criterion is based on the dilute
approximation, only points with \(p_{\rm sub}(T_n)<10^{-2}\) are included in the
fit.  For the comparison shown in Fig.~\ref{fig:fastcrit}, we also require a
finite \(\widehat f_\xi(T_n)\) and a nonzero kinetic volume fraction,
\(f_b(T_n)>10^{-12}\).  This avoids having the logarithmic fit dominated by
near-zero numerical values that already lie in the safe region.

As shown in Fig.~\ref{fig:fastcrit}, the fast criterion tracks the full kinetic
result over this selected sample.  A fit gives
\begin{equation}
\log_{10} f_b(T_n)
\simeq 0.20+1.13\log_{10}\fhat(T_n),
\label{eq:fit}
\end{equation}
with Pearson correlation \(r\simeq0.995\).  The percent level boundary
\(f_b(T_n)=10^{-2}\) corresponds to
\begin{equation}
\log_{10}\fhat(T_n)\simeq -1.95 .
\label{eq:cheapcriterion}
\end{equation}
Therefore \(\log_{10}\fhat(T_n)\gtrsim -1.95\) is a practical indication that the
homogeneous bounce calculation should be accompanied by a subcritical
prehistory calculation.  Far below this boundary, the homogeneous description
is self consistent.  Near or above it, the kinetic equation should be solved.
Points close to a class boundary can move when the source normalization or
erasure model is varied, as shown in Appendix~\ref{app:robustness}.  Thus the
numerical threshold is mildly model dependent.  Its role is as a
computationally inexpensive decision rule.

For a more accurate fast estimate, one may use two- or even multi-bin approximation.  In
these approaches, the subcritical radius range is divided into several bins. The abundance in each bin evolves separately. The formation and erasure
rates are evaluated bin by bin, while shrinking is treated as a flux from
larger-radius bins to smaller-radius bins. 


Finally, subcritical bubble may also have indirect observational consequences.  For electroweak baryogenesis, the impact is likely limited in the present scan, because sizable subcritical populations mainly occur in weak transitions $\phi_n/T_n < 1$, while
washout avoidance usually requires a stronger transition.  The situation can be different for filtered dark matter and PBH. In filtered dark matter scenarios, the relic abundance can depend on how particles encounter, transmit through, or become trapped by phase boundaries~\cite{Baker:2019ndr}. A pre-existing subcritical bubble can therefore change the early exposure of the plasma to broken vacuum regions by introducing additional stochastic phase boundaries. Particles reflected by an expanding critical wall may subsequently scatter from nearby subcritical bubbles, be redirected back toward the wall, or become temporarily confined between multiple interfaces. This modifies the effective filtering history and may enhance the final dark matter abundance.

Since PBH production is controlled by rare regions that stay in the symmetric phase for an unusually long time, a pre-existing subcritical bubble can disturb these would-be delayed patches by adding small broken vacuum regions inside them. These fluctuations may trigger earlier local conversion, prevent delayed regions from remaining fully in the symmetric phase, and change the probability of very late completion. As a result, the abundance, typical size, and clustering of PBH producing regions may be modified. 

For gravitational waves, subcritical bubbles are most prominent in the small
\(\alpha_n\) region, where the released vacuum energy is modest and the signal is expected to be dominated by sound waves rather than bubble
collisions.  Since the subcritical population turns on over a short interval
before \(T_n\), critical bubbles can be viewed as nucleating in a fluctuation background.  These fluctuations introduce length scales smaller than
the usual mean bubble separation, set by the correlation length and by the shrinking and erasure times.  The main sound wave peak may therefore remain controlled by the standard hydrodynamics, while the high frequency part of the spectrum may receive corrections from the smaller subcritical scales \cite{Bian:2026xdm}. 

The present analysis should be viewed as a fast consistency check rather than a
complete real time simulation.  It uses a one field benchmark potential, a
Gaussian ansatz, phenomenological formation and erasure rates, and a dilute
treatment of \(V_b^{(R)}/V_{\rm tot}\).  In realistic particle model scans, such as the real singlet
model or two Higgs doublet models,
the same logic can be applied with the full finite temperature effective potential, while points near dilute breakdown require a more precise evolution of the birth-death equation.

\section{Conclusion}
\label{sec:conclusion}
We studied subcritical bubble prehistory as a consistency test of the
homogeneous symmetric vacuum background assumed in standard 
nucleation picture.  By evolving a birth-death equation for the subcritical radius
distribution along a cosmological cooling history, we quantified the broken vacuum
volume fraction already present at the usual nucleation temperature.

Our parameter scan shows that the homogeneous nucleation background is most
vulnerable when the symmetric and broken vacua remain nearly degenerate at \(T_n\),
the barrier is low on the thermal scale, and the free energy of a
correlation-sized broken patch is small enough to overcome erasure.  The
largest subcritical bubble volume fractions are concentrated at small transition strengths,
typically \(\alpha_n\lesssim10^{-3}\), and modest order parameter values, while
the intermediate prehistory relevant band extends to larger \(\phi_n/T_n\) and
\(\alpha_n\).  The classification is therefore set by the combined free energy
landscape and cooling history rather than by a single number.

The same physical ingredients are captured by the fast criterion
\(\hat f_\xi(T_n)\), which estimates the volume fraction of a correlation-sized broken
phase fluctuation from its volume, formation rate, and lifetime.  The percent
level boundary in the scan corresponds to
\(\log_{10}\hat f_\xi(T_n)\simeq -1.95\).  Values far below this scale indicate
that the homogeneous bounce calculation is self consistent, while values near
or above it indicate that the critical bubble may form on a mixed background.
Finally, the mixed background induced by subcritical bubble may affect inhomogeneous nucleation, filtered dark matter, PBH, or the high
frequency part of the gravitational wave spectrum. 

\appendix

\section{Implicit update of the subcritical distribution}
\label{app:numerics}
The kinetic equation is stiff whenever the microscopic erasure time is shorter
than the Hubble cooling time.  We therefore use an implicit upwind update in
radius space.  For the shrink flow \(dR/dt=-v\), the discretized equation is
\begin{equation}
\frac{n_i^{m+1}-n_i^m}{\Delta t}
=v\frac{n_{i+1}^{m+1}-n_i^{m+1}}{\Delta R_i}
 +S_i^{m+1}-K_i^{m+1}n_i^{m+1},
\label{eq:upwind}
\end{equation}
where \(S_i=P_f\Gamma_{f\to t}(R_i,T)\).  In our kinetic setup
\(K_i=3H+aT+(4\pi R_i^3/3)\Gamma^{\rm sub}_{t\to f}\). 

For a sufficiently small timestep, we treat \(S_i\) and \(K_i\) as constant
over the interval \([t,t+\Delta t]\).  The resulting equation,
\begin{equation}
\frac{dn_i}{dt}=S_i-K_i n_i,
\end{equation}
can then be integrated analytically, giving
\begin{equation}
n_i(t+\Delta t)
=
n_i(t)e^{-K_i\Delta t}
+\frac{S_i}{K_i}\left(1-e^{-K_i\Delta t}\right).
\label{eq:relax}
\end{equation}

The empty volume factor is solved implicitly rather than taken from the previous
step.  Since the update is linear in \(S_i\), we first compute the no source
solution \(n_i^{(0)}\) and the full source solution \(n_i^{(1)}\), and then solve
\begin{equation}
P_f=\exp\left[-f_t\!\left(n^{(0)}+P_f[n^{(1)}-n^{(0)}]\right)\right]
\label{eq:implicitpoisson}
\end{equation}
by bisection.  This implicit Poisson step avoids spurious binary feedback when
the subcritical population relaxes much faster than the temperature changes.
The timestep follows from \(dt=-dT/(HT)\). 

\section{Numerical validation}
\subsection{Grid convergence}
\label{app:validation}
\begin{figure}
\includegraphics[width=\columnwidth]{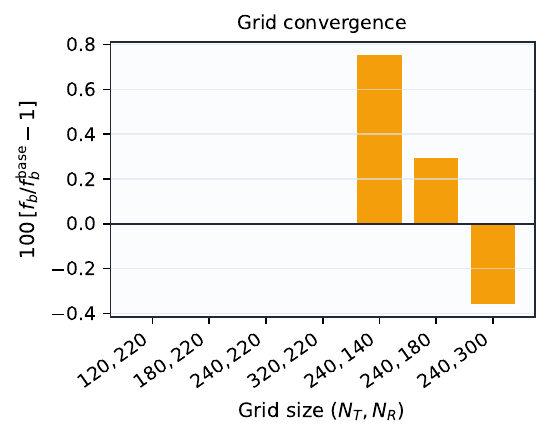}
\caption{\label{fig:validation}
Numerical checks about grid dependence of \(f_b(T_n)\) for a
prehistory relevant benchmark.}
\end{figure}

Fig.~\ref{fig:validation} summarizes two checks on the solver.  For the
benchmark point \((D,\lambda,r_c)=(0.35,0.21,0.40)\), varying the cooling grid between 120 and 320 points leaves \(f_b(T_n)\) unchanged at the precision shown.  The radius grid is more demanding because the direct volume integral weights the radius distribution by \(R^3\): the coarsest radius grid shifts the result at the ten percent level, while the finer grids reduce the change to a few percent. 

\subsection{Robustness of kinetic assumptions}
\label{app:robustness}
\begin{figure}
\includegraphics[width=\columnwidth]{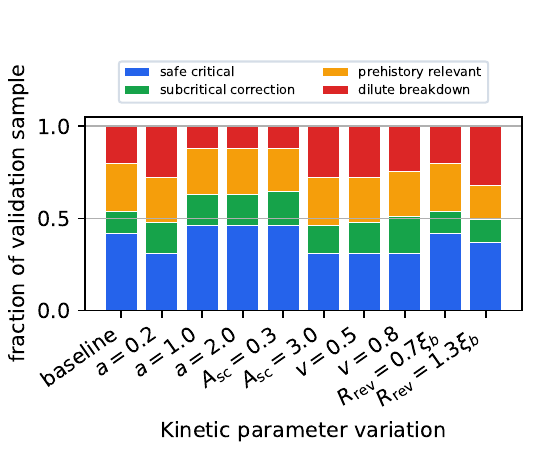}
\caption{\label{fig:robustness}
Class
fractions under variations of thermal erasure, source normalization, shrink
speed, and the reverse erasure radius cut.}
\end{figure}

The baseline scan uses \(a=0.5\), \(A_{\rm sc}=1\), \(v=1\), and the reverse
erasure cut \(R>\xi_b\).  Fig.~\ref{fig:robustness} tests the robustness of
these choices on a 65-point validation sample, selected from \(\alpha_n\)
quantiles and from points close to the classification thresholds.  For each
variation, we recompute the action curve and \(T_n\) before evolving the
subcritical kinetic equation.  Since \(v\) is a physical shrinkage speed, we
vary it only within the causal range \(0<v\le 1\), using \(v=0.5\) and
\(v=0.8\).

The resulting shifts follow the expected physical trends.  Increasing
\(A_{\rm sc}\), or weakening the erasure channels, increases the subcritical
fraction, while faster shrinkage within the causal range decreases it.
Changing the reverse erasure cut between \(0.7\xi_b\) and \(1.3\xi_b\) has a
smaller median effect than varying \(A_{\rm sc}\) or \(a\), although points near
the classification thresholds can still move between classes.

\begin{acknowledgments}
The numerical calculations in this work were carried out on the High-Performance Computing Platform at the Center for Theoretical Physics, Henan Normal University. This work was supported by the National Natural Science Foundation of China (Grant Nos. 12335005), the PI Research Fund of Henan Normal University (Grant No. 5101029470335), and the National Natural Science Foundation of Henan province(Grant No. 262300421233).
\end{acknowledgments}

\bibliographystyle{apsrev4-2}
\bibliographystyle{JHEP}
\bibliography{subcritical_pretransition_prd}

\end{document}